# Remote Sensing of Trace Element in Sea Salt Aerosol with Sensitivity Level of 10 pg/m$^3$

Yuezheng Wang [1,2#], Nan Zhang [1,2#], Jiayun Xue [1,2], Bingpeng Shang [1,2], Jiewei Guo [1,2], Zhi Zhang [1,3], Pengfei Qi [1,2,*], Lie Lin [1,3], Weiwei Liu [1,2*]

[1] Institute of Modern Optics, Eye Institute, Nankai University, Tianjin 300350, China;
[2] Tianjin Key Laboratory of Micro-scale Optical Information Science and Technology, Tianjin 300350, China;
[3] Tianjin Key Laboratory of Optoelectronic Sensor and Sensing Network Technology, Tianjin 300350, China

**Abstract**: Sea salt aerosols composed mainly of micrometer-sized sodium chloride particles not only pose a potential threat to human health and traffic safety, but also directly affect climate prediction. The long-range and high-precision sensing of sea salt aerosols remains a challenge for existing composition analysis methods. As the development of ultrashort laser technology, femtosecond laser filamentation provides a new opportunity for molecular remote sensing in complex environments. However, the accuracy at long-distance of this method is still hard to meet the demand (<10 ng/m$^3$) for the remote aerosol monitoring. To solve this problem, we built a remote detection system for sea salt aerosol fluorescence spectroscopy and obtained a very high system sensitivity by introducing a terawatt-class high-performance femtosecond laser and optimizing the filament and aerosol interaction length. The system achieves a Na$^+$ detection limit of 0.015 ng/m$^3$ at a detection distance of 30 m, and 0.006 ng/m$^3$ when supplemented with a deep processing learning algorithm.

Sea spray aerosol, also known as sea salt aerosol (SSA), refers to a kind of inorganic salt aerosol with sodium chloride as the main component produced on the ocean surface driven by wind[1,2]. SSA has the characteristic of hygroscopic growth[3], which makes it possible to indirectly impact the climate system by acting as cloud condensation nuclei (CCN) and ice nuclei (IN)[4], and altering the microphysics and radiative properties of clouds (indirect effects). Additionally, aerosols can modify cloudiness by heating the atmosphere where clouds reside (semi-direct effect), and reduce the reflectivity of snow, land, and sea ice through deposition and melting. Therefore, in predicting future climate change, aerosol remains one of the largest sources of uncertainty[5,6,7,8]. More notably, when these micrometer-sized particles come into contact with living organisms, they can easily cause harm to the respiratory and cardiovascular systems[9], and the atmospheric turbidity and visibility deterioration caused by the mixing of aerosol particles are easy to cause traffic jams and traffic accidents[10]. In summary, the quantitative and qualitative detection of sea salt aerosols is of great significance for climate prediction as well as the prevention and evaluation of atmospheric environmental pollution[5, 11].

There are many factors affecting the generation and distribution of sea salt aerosols[2]. Currently, climatologists mainly simulate sea salt aerosols based on wind speed and sea surface temperature[12, 13], and the simulation results of different models vary greatly[14], so there is an urgent need for compositional analysis tools with real-time monitoring and remote sensing capabilities. In addition, the concentration of marine aerosols in the atmosphere is usually less than 10 (ng/m$^3$)[11], which requires methods with high sensitivity[15]. Although the elemental detection methods represented by X-ray fluorescence (XRF) have high sensitivity and can reach the level of μg/m$^3$ to ng/m$^3$, they do not have remote sensing capability[16, 17].

Laser-induced plasma spectroscopy provides a versatile method for trace element measurements[18, 19]. The principle is that ultra-short pulse laser is focused on the surface of the sample, when the laser irradiance exceeds the breakdown threshold of the sample, the particles (atoms, molecules) in the laser ablation zone are ionized to form a plasma, and the excited particles transition from high energy levels to low energy levels and emit characteristic spectral lines. Among them, the wavelength and intensity of atomic emission spectrum represent the type and content of the measured substance respectively[20]. Laser-induced plasma spectrum has a significant impact in a wide range of areas[21, 22, 23, 24]. Unfortunately, the current long-range detection accuracy of laser induced plasma spectroscopy is not up to the requirement of atmospheric aerosol detection. The problem is that the power density attenuation value of femtosecond laser increases with increasing penetration depth after crossing clouds, raindrops and other obstacles[25]. And one of the important factors affecting the sensitivity of spectral detection is the power density[26].

Femtosecond laser pulses can overcome natural diffraction effects when transmitted in transparent media such as air and glass[27]. Due to the dynamic balance between self-focusing induced by the optical Kerr effect and defocusing from the self-generated plasma in the self-focal region[28], the laser pulses converge in a plasma channel (filament) with a diameter of about 100 μm, thus enabling long-range transmission (0-10 km)[29]. The light intensity in the "filament" is up to $10^{13} \sim 10^{14}$ W/cm$^2$, which allows direct excitation of most substances to form plasma[30, 31].

In this paper, a remote trace element spectroscopy detection system is designed. The system consists of a terawatt-class high-performance femtosecond laser (1030 nm, 200 mJ, 100 Hz), a long-range femtosecond laser filamentation optical path that integrates "receiving" and "transmitting", and a spectrometer. Using this system, we obtained a detection limit of 0.006 ng/m$^3$ for sodium ions in sea salt aerosol at a detection distance of 30 m. This index is the latest record in the world for the detection of sodium ions in sea salt aerosols by spectroscopic techniques, thus also demonstrating the potential impact of the method in the field of atmospheric remote sensing.

## Results and Discussions

**Filament-induced plasma spectroscopy.** Previous researchers have conducted a series of studies on aerosols using filament-induced plasma technology[32]. The first issue discussed was the long-range interaction between the filament with the aerosol, namely whether the filament can penetrate obstacles such as clouds or raindrops in a complex atmospheric environment. Analysis has shown that the light intensity clamping effect results in only 10% of the energy being concentrated inside the filament, with the remaining energy forming a background energy reservoir around the filament[33]. The energy stored in the background energy reservoir can be replenished to the filament, allowing for long-range transmission.

Now the critical issue to be investigated is how to improve the detection sensitivity at long distance for the filament induced plasma spectroscopy technique[34]. For example, a distinctive feature of the femtosecond laser filamentation process is the spectrum broadening, which results from the combination of various nonlinear effects such as self-phase modulation, self-steepening, and dielectric ionization[33, 35]. The supercontinuum spectrum covers the wavelength range from ultraviolet to infrared, which happens to mask the fluorescence emission wavelengths of some metal cations and adversely affects both the qualitative and quantitative analysis of material composition. This issue can be eliminated through polarization gate switching technology[36] or time-resolved techniques[37].

Another issue worth analyzing is the placement of the laser remote-induced plasma spectroscopy collection device. When femtosecond laser filaments propagate through the atmosphere, the method of using air as a gain medium to produce coherent radiation is called air lasing[38], and its physical mechanisms include amplified spontaneous emission (ASE) and amplified stimulated emission[39]. Among them, ASE enhances the backward fluorescence signal radiated by the formation of plasma after the filament excites the sample and binds the originally spatially distributed isotropic fluorescence signal to a small angular range near the filament axis[40], directed toward the emission end and facing the observation device. This interesting phenomenon fits well with the application requirements of atmospheric remote sensing, where both transmitting and receiving at the same time in the ground segment.

**Sea salt aerosol detection system.** In this paper, we have established a remote sensing system of sea salt aerosol fluorescence spectrum, as shown in Figure 1(a). Figure 1(d) shows the pulse duration of the incident laser, measured using the FROG from Femtoeasy, with a FWHM of 496.9 fs. The incident femtosecond laser （1030 nm, 200 mJ, 100 Hz） is first reflected by the Mirror 1 and then focused by a combined lens system (concave lens, concave mirror). The onset of femtosecond laser filamentation was controlled at a distance of approximately 30 m from the concave mirror by varying the relative distance between the concave lens (F = -15 cm; R = 2.54 cm) and the concave mirror (F = 200 cm; R = 20.7 cm). According to the reversible optical path principle, the filament interacts with the aerosol, generating a backward fluorescent signal of plasma radiation, which is collected and focused by a concave mirror. We placed a fiber behind Mirror 2 (dichroic mirror: @1030 nm reflectance > 99 %, @589 nm transmittance > 94.9 %), the fiber transmitted the received fluorescence signal to a grating spectrometer (Omni-λ300, Zolix) with slit, and the spectral information was output by an Istar-sCMOS camera and recorded to computer. The gate delay, gate width and exposure time of Istar-sCMOS are set to 220 ns, 1000 ns and 30 s, respectively, to obtain spectral images with high signal-to-noise ratio.

In addition, it is difficult to achieve large diameter, light mass and low cost for long-range femtosecond laser filamentation systems composed of transmissive optical elements (convex lenses), while the reflected femtosecond laser filamentation system used in this paper inevitably introduces aberration. Conventional ways of eliminating aberrations include replacing concave lenses with anamorphic mirrors, spatial light modulators (SLM)[41], and free-form phase plates[42], all of which cannot be used at extremely high power. In this paper, an asymmetric incidence is used to break the symmetry of the spot distribution on the concave mirror, which can effectively compensate the wavefront aberration caused by the off-axis system[43].

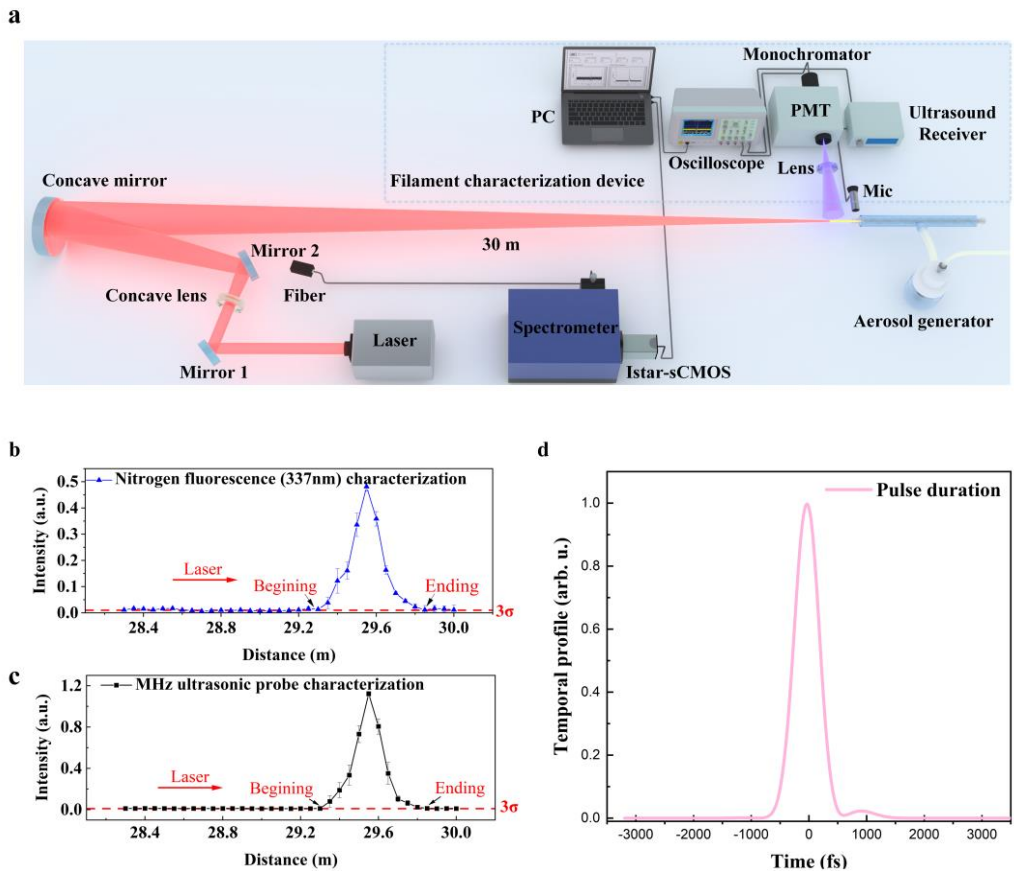

**Fig. 1 a** Schematic diagram of sea salt aerosol spectral detection system based on femtosecond laser filamentation. Mirror1: plane mirror, Mirror2: dichroic mirror, @1030 nm reflectance > 99 %, @ 589 nm transmittance > 94.9 %, Concave lens (F = -15 cm; R = 2.54 cm), Concave mirror (F = 200 cm; R = 20.7 cm), Lens: Convex lens (F = 5 cm), Mic: Microphone. **b** The fluorescence signal at 337 nm with nitrogen was used to characterize the filament, and the blue line in the graph represents the amplitude value of the fluorescence signal at different locations along the filament axis. **c** Characterization of filament by ultrasound signal. The black dots in the graph are the amplitude values of the ultrasonic signals at different locations along the filament axis, and the red line represents three times the standard deviation of the ambient noise(3σ). **d** Pulse duration was measured with a FROG from Femtoeasy (FWHM:496.9 fs).

Figure 1(b) and Figure 1(c) shows the starting position and length of the filament characterized by the ultrasound signal and the nitrogen fluorescence signal, respectively. We placed microphone (V306, Olympus) on the side of the filament to collect the acoustic signal excited by the filament. The acoustic signals were amplified by an ultrasonic pulse receiver (5072PR, Olympus) and displayed on a digital fluorescent oscilloscope (DPO3034, Tektronix Inc), which was recorded by a computer. Meanwhile, a convex lens with a focal length of 5 cm, a monochromator (WGD-100, Gang Dong Sci. & Tech. Co. Ltd.), and a PMT (H11902, Hamamatsu) were placed to collect the fluorescence signal of nitrogen (337 nm) excited by the filament in the environment. By moving the ultrasonic probe (MHz) and fluorescence receiver device point by point (5 cm steps) along the axial direction of the filament using a manual displacement table, the spatial position distribution and axial relative intensity distribution of the femtosecond laser filament formation in this experiment can be recorded in detail. After a simple data processing, the starting point of the filament is 29.30 m from the concave mirror and the ending point is 29.85 m from the concave mirror, giving a length of 0.55 m for the complete filament.

It is worth noting that the detection sensitivity of the aerosol signal is related to the fluorescence collection efficiency of the spectrometer, therefore the following optimization were taken to ensure that the fiber is in the optimum position. A small bulb was placed at the focal point of the femtosecond laser to simulate the fluorescence signal generated backwards after aerosol excitation. According to the principle of reversibility of the optical path, the light from the small bulb is collected by the concave mirror and converges to the fiber on the back of the Mirror 2. The position of the fiber is optimized point by point by means of a precision manual 3D translation table to obtain the maximum signal value as the optimum position of the fiber.

**Remote Sensing of $Na^+$ in Sea Salt Aerosol.** When using femtosecond laser filaments as a pump source to excite sea salt aerosols, the reverse fluorescence signal is affected by forward amplified spontaneous emission (ASE) and Mie scattering. Based on numerical simulations of the nonlinear wave equation and Mie scattering, the changes in aerosol backward fluorescence radiation can be analyzed. First, the propagation of forward fluorescence can be expressed as equation (1)[27].

$$-2ik\frac{\partial A}{\partial z} = \Delta_\perp A + 2\frac{k^2}{n_0}n_{nl}A - ik\alpha A \quad (1)$$

In the equation (1), $A$ represents the fluorescence distribution, $k$ is the wavenumber corresponding to $\lambda$ = 1030 nm, $n_{nl}$ is the refractive index change caused by the light filament during fluorescence propagation, and $\alpha$ is the absorption caused by the aerosol. The next step is to analyze the spatial distribution of Mie scattering signal intensity, that is, to calculate the backscattered fluorescence radiation of aerosols, which can be represented by equation (2) and (3)[44].

$$I(\theta) = I_0 S_j(\theta) / (k^2 d^2) \quad (2)$$

$$|S_j(\theta)|^2 = \left| \sum_n \frac{2n+1}{n(n+1)} (a_n \pi_n + b_n \tau_n) \right|^2 \quad (3)$$

Assuming that the aerosol is a symmetrically distributed spherical particle, $I(\theta)$ represent the scattering intensity measured at a distance d from the scattering sphere, and $I_0$ represents the fluorescence signal intensity obtained after the interaction between the aerosol and the filament. Thus, according to the theory of geometric relationship of spatial collection angles, the spatially integrated intensity of the backward fluorescence signal collected by the concave mirror is calculated, as shown by the red curve in Figure 2c.

In addition, we also tested the optimal interaction length between the filament and the aerosol in the experiment to obtain the best $Na^+$ fluorescence spectrum. The experimental process is as follows: By moving the glass tube filled with NaCl aerosol particles ($Na^+$ mass concentration of 0.3 mg/m$^3$) horizontally to the left point by point (5 cm step), the interaction length between the filament propagating to the right and the NaCl aerosol gradually increases, as shown in Figure 2(a). The spectral intensity of $Na^+$ at 589 nm increases and then decreases, and the intensity of the fluorescence signal excited by $Na^+$ in sea salt aerosol is maximum when the aerosol contact is close to completely "engulfing" the filament (interaction length 45 cm, total filament length 55 cm), as shown in Figure 2(b). The experimental results represented by the blue dots in Figure 2c agree well with the theoretical calculations represented by the red curve. Analysis indicates that the gain length increases as the glass tube moves towards the filament, allowing fluorescence photons to continuously generate and amplify in the gain medium ($Na^+$). However, the femtosecond laser pulse suffers from high transmission loss in the aerosol before forming the filament. When the interaction length approaches or exceeds the length of the filament, the energy loss caused by Mie scattering results in a weakened filament. Thus, the gain effect is limited by the energy loss caused by Mie scattering.

Next, we performed the mass concentration detection limit analysis of sea salt aerosol and tested the fluorescence spectra of $Na^+$ at each concentration, and the test results are shown in Figure 3a.

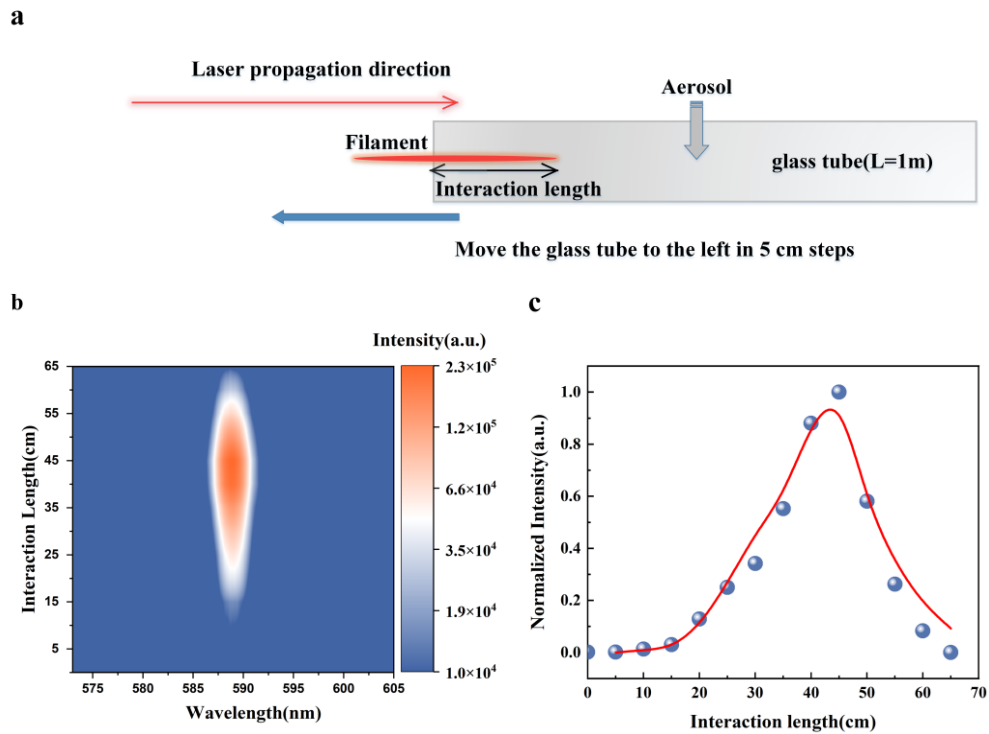

**Fig. 2** Characterization of interaction length between sea salt aerosol and filament. **a** Diagram illustrating the variation of the interaction length between the laser filament and aerosol. **b** Fluorescence spectrum (pseudo color plot) of $Na^+$ under different interaction lengths of aerosol and filament. **c** Integrated intensity values (Normalization was performed) of $Na^+$ fluorescence spectral lines at 589 nm for different interaction lengths of aerosols and filaments. The blue data points represent the experimental measurements, while the red curve shows the data calculated using the nonlinear equation and Mie scattering theory.

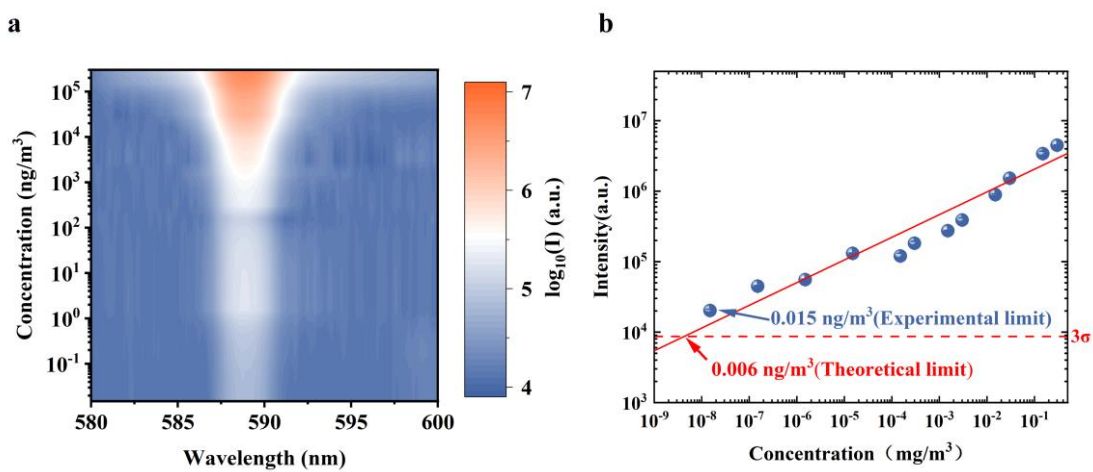

**Fig. 3** Measurement of detection limit of sea salt aerosol. **a** Fluorescence spectrum (pseudo color plot) of $Na^+$ under different concentrations of sea salt aerosol. **b** The plot of the integrated fluorescence intensity of $Na^+$ (589 nm) as a function of sea salt aerosol concentration. The blue dots represent the experimental measurements, and the intersection of the fitted line with the 3σ threshold represents theoretical prediction value.

The fluorescence spectrum of Na$^+$ at 589 nm is clearly visible for each concentration of sea salt aerosol excited by the filament, and the concentration is positively correlated with the peak intensity of the spectral line. To show the detection limit of sea salt aerosol more clearly, the integrated intensity corresponding to the spectral line of Na$^+$ at 589 nm was taken to obtain Figure 3(b). The intensity of the fluorescence signal generated by the filament-induced aerosol can be expressed according to the lidar equation as follows[45]:

$$E(\lambda, R) = E_L K_0(\lambda) T(\lambda, \lambda_L R) \xi(R) \frac{A_0}{R^2} N(R) \frac{\sigma^F(\lambda_L, \lambda)}{4\pi} \frac{c\tau_d}{2} \quad (4)$$

Where $E_L$ represents the energy density in the filament as a function of the propagation distance R. $K_0(\lambda)$ is the spectral collection efficiency of the acquisition system. $T(\lambda, \lambda_L R)$ is the atmosphere transmission factor and $\xi(R)$ is the geometrical overlapping factor which dependent on the geometrical dimensions of the telescope and the divergence of the laser beam (filaments). $N(R)$ is the number density of the excited molecule number density. Clearly, the relationship between the aerosol particle concentration and the integrated intensity of the fluorescence signal is obtained in direct proportion. The optimal detection limit of sea salt aerosol was experimentally measured to be 0.015 ng/m$^3$. By linear fitting, we obtained the linear function of concentration versus spectral integrated intensity, and the intersection of this function with 3σ (three times the standard deviation of the background noise) is the theoretical optimal detection limit of 0.006 ng/m$^3$ for the detection of sea salt aerosols by our built system.

In conclusion, a sea salt aerosol fluorescence spectroscopy detection system that enables long-range and real-time measurements is built using femtosecond laser filamentation as the pump source. After a series of preparations, we experimentally obtained the optimal Na$^+$ detection limit of 0.015 ng/m$^3$ at a detection distance of 30 m. By using the Na$^+$ mass concentration versus the intensity of the characteristic spectral lines linear relationship between the Na$^+$ mass concentration and the intensity of the characteristic spectral lines, the Na$^+$ detection limit of 0.006 ng/m$^3$ at a distance of 30 m was theoretically predicted for this system. This work provides an effective guide to the experimental methods and data processing for remote trace detection in atmospheric aerosols by high power density femtosecond lasers.

# Methods

**Ultrasonic signal characterization of filament.** The principle of characterizing filaments by acoustic signals is as follows: When the femtosecond laser propagates in the air and converges into a filament, the free electrons formed by ionization of air molecules are emitted with high kinetic energy (several eV, corresponding to an initial electron temperature of the order of $10^4 \sim 10^5 K$) to transfer energy in the form of elastic and inelastic collisions with the surrounding medium and produce an expanding column of hot gas after the plasma compound, thus emitting an acoustic wave that can be detected by a microphone[46]. The intensity of the acoustic wave is proportional to the laser energy absorbed in the filament and therefore proportional to the initial free electron density. Such sonographic detection has been demonstrated to be a simple and sensitive method of determination of the spatial extent and longitudinal plasma profile of a filament by translating a microphone along a filament.

**Fluorescence signal characterization of filament.** The principle of characterizing filaments by nitrogen fluorescence signals is as follows[47]: The volume fraction of nitrogen in air is about 78.08%, and the femtosecond laser is easy to ionize nitrogen molecules into the ground and excited states when filamentation in air. The first negative band gap radiates fluorescence through $\left( B^2\Sigma_u^+ \to X^2\Sigma_g^+ \right)$, and the second positive band system radiates fluorescence through $\left( C^3\Pi_u \to B^3\Pi_g \right)$ such a collisional complexation and relaxation process in the wavelength of 337 nm. Since the fluorescence emission is a result of photon/tunnel ionization, the intensity of the fluorescence is directly related to the laser intensity distribution and therefore to the plasma density within the filament, again allowing the spatial extent of the filament and the longitudinal plasma profile to be determined[43].

**Sea salt aerosol generator.** In this thesis, sea salt aerosols are generated by an aerosol generator (HRH-WAG3, Beijing HuiRongHe Technology Co., Ltd, China), which is capable of uniformly and stably producing aerosol particles with a median mass particle size between 1-3 μm. In the previous work[48], we calibrated the concentration of aerosol by extinction value method. In this experiment, the mass concentration of $Na^+$ particles $1.5 \times 10^{-7}$ to 0.3 mg/m$^3$ in the aerosol particle system were achieved by using NaCl solution with a mass fraction ranging from $5\times10^{-7}$ to 1 percent in a 1m long glass tube.

**Acknowledgements**

This research was funded by the National Key Research and Development Program of China (2018YFB0504400) and Fundamental Research Funds for the Central Universities (63223052).

**Author contributions**

Yuezheng Wang: Data curation, Writing – original draft, Software.

Nan Zhang: Conceptualization, Methodology, Writing – review & editing.

Jiewei Guo: Visualization.

Zhi Zhang: Investigation.

Bingpeng Shang: Resources.

Lie Lin: Resources.

Pengfei Qi: Writing – original draft, Supervision.

Weiwei Liu: Conceptualization, Methodology, Validation, Supervision.

**Competing interests**

The authors declare no competing financial interests.